\documentclass[aps,prl,showpacs,twocolumn,groupedaddress,amssymb]{revtex4}

\usepackage{graphicx}
\usepackage{dcolumn}
\usepackage{bm}

\begin{document}
 

\title{Helical relativistic electron beam and THz radiation}

\author{S. Son}
\affiliation{18 Caleb Lane, Princeton, NJ 08540}
\author{Sung Joon Moon}
\affiliation{28 Benjamin Rush Lane, Princeton, NJ 08540}
\date{\today}

\begin{abstract}
A THz laser generation utilizing a helical relativistic electron beam propagating
through a strong magnetic field is discussed.
The initial amplification rate in this scheme is much stronger than that in
the conventional free electron laser.
A magnetic field of the order of Tesla can yield a radiation in the range of 0.5 to 3 THz,
corresponding to the total energy of mJ and the duration of tens of pico-second,
or the temporal power of the order of GW.
\end{abstract}

\maketitle

A THz electromagnetic (E\&M) wave has a range of practical 
applications~\cite{diagnostic, radar, seigel3, security}. 
In particular, the light wave of the frequency of 1 to 10 THz 
is under increased attention~\cite{siegel, siegel2}. 
Around the frequency range of 100 GHz,  
there exist appropriate technologies such as gyrotron~\cite{gyrotron, gyrotron2, gyrotron3}.
However, these technologies cannot be extended to the range above a few hundred GHz,
due to the well-known scaling problem~\cite{booske}. 
Other technologies such as the quantum cascade laser~\cite{qlaser, qlaser2}
and the free electron laser~\cite{colson} 
have their own limitations in generating an intense E\&M wave. 
There have been preliminary proposals for generating a THz radiation based on
the recent advances in the intense visible laser~\cite{Fisch, sonlandau, sonbackward},
in the context of the inertial confinement fusion~\cite{tabak, sonpla, sonprl}. 

\begin{figure}
\scalebox{0.5}{
\includegraphics{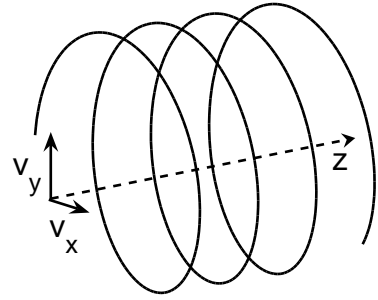}}
\caption{\label{fig1}
The helical velocity structure of the relativistic electron beam propagating
with a constant velocity along a slightly skewed direction to the $z$-axis.
This schematic diagram is not drawn to scale.}
\end{figure}

In this paper, we propose a scheme to generate a THz radiation,
where the energy is extracted from the perpendicular kinetic energy of an relativistic electron beam,
in the presence of a strong magnetic field. 
In this scheme,
a relativistic electron beam gets launched to a slightly skewed direction with respect to the magnetic field.
The electrons gyrate around the magnetic field and the perpendicular velocity of the electron
exhibits a periodic structure (Fig~\ref{fig1}). 
When a certain resonance condition is satisfied,
a specific E\&M THz wave becomes amplified. 
In particular, the rate at which the electron energy is extracted
is proportional to the electric field strength, as opposed to the energy intensity as
in the conventional free electron laser (FEL). 
This difference leads to a much more explosive amplification
compared to the conventional FEL. 
The goal of this paper is to  estimate the amplification efficiency.

We start by describing the motion of the helical electrons and the FEL amplification,
and then describe a new scheme enabling an explosive amplification.
A relativistic electron moving in the presence of the magnetic field is described by
\begin{equation} 
m_e\frac{ d \gamma_0 \mathbf{v}}{dt}= -e \frac{\mathbf{v}}{c} \times  \mathbf{B} \mathrm{,} \label{eq:b}
\end{equation}
where  $m_e $ is the electron mass, $\mathbf{v}$ is the electron velocity, 
$\gamma_0^{-2} = 1 - (v_{0z}^2 + v_p^2)/c^2$ is the relativistic factor, $v_{0z}$ ($v_{p}$ is
the parallel (perpendicular) velocity relative to the $z$-axis,
and the magnetic field is given as $ \mathbf{B} = B_0 \hat{z}$. 
The solution is   $v^{(0)}_{z}(t) = v_{0z}$, $v^{(0)}_{x}(t) = v_p \cos(\omega_{ce} t +\phi_0)$, and $v^{(0)}_{y}(t) = -v_p \sin(\omega_{ce} t +\phi_0)$, where $\omega_{ce} = eB_0/\gamma_0 m_e c $.
Consider a linearly-polarized E\&M wave propagating along the $z$-direction
so that  $E_x(z,t) = E_1 \cos(kz - ckt)$, 
$E_y=E_z=0$, $B_y(z,t) =  E_1\cos(kz - ckt) $, and $B_x = B_z = 0$.
The perturbed motion of the electron is
\begin{eqnarray} 
m_e\left(\frac{ d \gamma_0 \mathbf{v}^{(1)}}{dt}
 + \frac{d\gamma_1 \mathbf{v}^{(0)}}{dt}\right)  
&=& - e \frac{\mathbf{v}^{(1)}}{c} \times  \mathbf{B}_0 \nonumber \\ \nonumber \\ 
&=&-e\left[ \mathbf{E}_1 + \frac{\mathbf{v^{(0)}}}{c}
 \times  \mathbf{B}_1 \right] \mathrm{,} \nonumber \\ \nonumber 
\end{eqnarray}
where $\mathbf{E}_1 = E_1\cos(kz - ckt) \hat{x}$, 
and $\mathbf{B}_1  = E_1 \cos(kz - ckt)  \hat{y}$ is the magnetic field of the E\&M wave. 
Consider the first  case 
when  $v^{(0)}_{z} \gg v^{(0)}_{x} $ and  $v^{(0)}_{z} \gg v^{(0)}_{y} $
so that the terms of $v^{(0)}_{x} $ and $v^{(0)}_{y}$ in the linearized first order equation,
compared to $v^{(0)}_{z} = v_{0z}$, can be ignored. 
The linearized equation is given as 
\begin{eqnarray} 
 m_e\gamma_0\frac{dv^{(1)}_{x}}{dt} 
 = -eE_1 \left(1 - \frac{v_{0z}}{c}\right) - eB_0 \frac{v^{(1)}_{y}}{c} \mathrm{,} \nonumber \\
m_e \left(\gamma_0\frac{dv^{(1)}_{z}}{dt} + \gamma_0^3 \frac{v_{0z}^2}{c^2} 
 \frac{dv^{(1)}_{z}}{dt}\right) = -eE_1 \frac{v^{(0)}_{x}}{c} \nonumber \mathrm{,} \\
m_e\gamma_0 \frac{dv^{(1)}_{y}}{dt}  = eB_0 \frac{v^{(1)}_{x}}{c}  \mathrm{,} \nonumber \\ \nonumber 
\end{eqnarray} 
where $E_1 = E_1 \cos(k z - ck t) $ 
and $v^{(0)}_{x}(t) = v_p \cos(\omega_{ce}t +\phi_0)$, $v^{(0)}_{z} = v_{0z} = \mathrm{const} $. 
The solution in the perpendicular direction can be obtained by using the complex coordinate $v_p(t) = v^{(1)}_{x} + i v^{(1)}_{y}$
such that  $v_p(t) $ is the solution of the following equation: 
\begin{equation} 
\frac{dv_{p}}{dt} + i\omega_{ce} v_p =  -\frac{eE_1}{ \gamma_0 m_e} \left[1- \frac{v_{0z}}{ c} \right] \mathrm{.} \nonumber 
\end{equation}
Then, the electron energy loss rate by the E\&M wave  per unit volume is
\begin{eqnarray} 
 \frac{d\epsilon}{dt} &=& n_e m_ec^2 \langle \frac{d\gamma}{ d t} \rangle \nonumber \\
& =&  n_e m_e \gamma^2_0  
\langle \frac{d\mathbf{v}^{(1)}}{ dt}\cdot \mathbf{v}^{(0)} \rangle   \nonumber \\
&=& n_e \gamma_0^3 \left[ 
\frac{ \frac{v_{0z}}{c}}{ \gamma_0 + \gamma_0^3  \frac{(v_{0z})^2}{c^2}}
 + \frac{(1-\frac{v_{0z} }{c})}{\gamma_0} \right]
 \langle eE_1 v^{(0)}_{x} \rangle \mathrm{,} \nonumber \\ \label{eq:b3}
\end{eqnarray} 
where 
$\langle \rangle$ is the ensemble average over the phase $\phi_0$, and  $eE_1 v_{0x}$ is,  from Eq.~(\ref{eq:b}), given  as 
\begin{equation} 
   eE_1 v_{0x} =  \frac{e E_1 v_p}{c} \cos( k z - ck t + \omega_{ce} t + \phi_0)\mathrm{.} \label{eq:det}
\end{equation} 
Consider the second case $v_{0z} \ll v_{0x} $ and  $v_{0z} \ll v_{0y}$,
where the computation is more complicated, in the absence of a closed-form solution. 
Retaining only the resonance term, the energy loss rate becomes
\begin{equation} 
 \frac{d\epsilon}{dt} =  n_e \gamma_0^3 \left[ 
\frac{1}{ \gamma_0 + \gamma_0^3  \frac{v_{p}^2}{2c^2}} \right]
 \langle eE_1 v^{(0)}_{x} \rangle \mathrm{.} \label{eq:b4}
\end{equation}
The resonance condition 
for the FEL amplification  is  $ v_{0z} k - ck + \omega_{ce} = 0$,
or $ k = \omega_{ce} / ( c - v_{0z}) $.
With the resonance condition being satisfied, 
the ensemble average of the energy loss or gain cancels out
in the first order of $E_1$ if the distribution over the phase angle $\phi_0$ is uniform.
The ensemble average in the second order of $E_1$ provides 
the conventional FEL amplification.

Now, let us dicuss the difference between the conventional electron beam
and the helical beam.
We note that there exists circumstances where the ensemble average
$\langle eE_1 v^{(0)}_{x} \rangle $  
does not cancel out in the first order of $E_1$.
Consider the time slice at $t=0$, where
the helical structure of the electron velocity is given as in Fig.~\ref{fig1}: 
\begin{eqnarray} 
v_{0x}(t=0,z) &=& v_p \cos(k_h z) \nonumber \\ \nonumber \\
v_{0y}(t=0,z) &=& -v_p \sin(k_h z) \nonumber \\ \nonumber \\
v_{0z}(t=0,z) &=& v_{0z} \nonumber \mathrm{,} 
\end{eqnarray}  
where $k_h = \omega_{ce} / v_{0z} $ is the helix wave vector. 
This helical structure, formed by the electron gun, has zero phase velocity in the laboratory frame or is a static wave. 
An electron initially ($t=0$) located at $z=z_0$ evolves as 
\begin{eqnarray} 
v_{0x}(t) &=& v_p \cos(k_h z_0 + \omega_{ce} t) \nonumber \\ \nonumber \\
v_{0y}(t) &=& -v_p \sin(k_h z_0 + \omega_{ce} t) \nonumber \\ \nonumber \\
v_{0z}(t) &=& v_{0z} \nonumber \mathrm{.} \
\end{eqnarray}
Since $E_1(z,t) = E_1 \cos(kz -ckt) $, the ensemble average over the electrons, 
$\langle EV \rangle  = e\langle E_z (v_{0x} + i v_{0y}) \rangle $, is given as
\begin{eqnarray} 
\langle EV \rangle = \int E_1 v_p \cos\left( k(v_{0z}t + z_0) -ckt +\omega_{ce}t - k_h z_0\right)  dz_0 \nonumber \\ 
= \int E_1 v_p \cos\left( (kv_{0z}-ck + \omega_{ce} )t +  (k_h+k) z_0\right)  dz_0 \nonumber
\mathrm{.}  
\end{eqnarray} 
The phase angle $\phi_0(z) $ is given as $\phi_0(z) = (k_h + k)z$.
While the ensemble average $\langle \rangle $ over the entire beam does cancel out,
it does not locally for a fixed value of $z$.
The electrons of the same phase, located in the range $ z_0 -\delta z < z < z_0 + \delta z $  where $\delta = \pi/2(k+k_H)$,
contribute coherently to $\langle EV \rangle$ so that 
the local E\&M wave is amplified or damped by the these 
coherently phased electrons.
The E\&M wave that itnitially gets amplified by the coherent electrons
gets damped by different but coherently phased electrons as it propagates. 
The frequency at which the E\&M wave experiences the change
between the amplification and the damping is estimated to be
$\Omega \cong 1/\delta t $, where $\delta t (c-v_0) = \delta z $.
Since $(c-v_{0z})k =\omega_{ce}$, $\delta t$ can be estimate as  $1/\omega_{ce}$ so that $\Omega \cong \omega_{ce}$.
This local amplification is in contrast with the electrons
with random phases for fixed $z=z_0$. 

The above argument suggests that
there exists a local amplification mechanism for the helical plasma,
that could be used as a THz generation.
Denoting the relativistic factors $\gamma_m = (1-v_{0z}^2/c^2)^{-1/2}$ and $\gamma = (1-v^2/c^2)^{-1/2}$,
the frequency of the amplified wave can be drived from the resonance condition $ k(c-v_{0z}) = \omega_{ce}$ as   $2\gamma_m^2 \omega_{ce}$.
Consider the case $\gamma_m = 7$,  $\gamma = 10$ and $B_0 = 1  \ \mathrm{T} $,
which correspond to $ 2 \gamma_m^2 \omega_{ce}=   300 \ \mathrm{GHz}$.
Consider another 
when $\gamma_m = 30$,  $\gamma = 40$ and $B_0 = 1 \ \mathrm{T} $,
$2 \gamma_m^2\omega_{ce} = 1.4 \ \mathrm{THz}$. 

Let us estimate the E\&M wave growth rate for the amplification.
For simplicity, let us use the reference frame where we
move with the electron beam with the same velocity in the $z$-direction.
Assume that $\gamma_m > 1$, $v_{0zm}\ll v_{0xm} $ and $v_{0zm} \ll  v_{0ym}$. 
If $v_{0x}$ ($v_{0y}$) is the perpendicular velocity in the laboratory frame, 
$v_{0xm} = \gamma_m  v_{0x} $ ($v_{0ym} = \gamma_m  v_{0y}$)
is the perpendicular velocity in the moving frame. 
If the electron density in the laboratory frame is $n_{e} $, then it is 
$n_{em} = n_{e} / \gamma_m$ in the moving frame. 
The electron energy loss rate in the moving frame given in Eq.~(\ref{eq:b4}) is 
\begin{equation} 
\frac{d\epsilon}{dt} \cong \alpha n_{em} \langle eE_1 v_{0xm} \rangle \mathrm{,}
\end{equation} 
where $\alpha$ is a constant of order of 1. 
By considering the local E\&M wave and the local amplification,
we obtain from  $d\epsilon / dt = (1/8\pi) (d E_1^2 / dt) $ that
\begin{equation}
\frac{dE_1}{dt} \cong \alpha 4 \pi e n_{em} v_{0xm} \cos(\Omega t) \mathrm{,} \label{eq:vom} 
 \end{equation}
where $\Omega \cong eB_0/m_e c $ is in the moving frame.
Eq.~(\ref{eq:vom}) shows that
the initial growth rate, $(dE_1/dt)/E_1$,  is infinite.
During the time duration of $1/\Omega$,  $E_1$
grows to  $E_1(T) = \alpha n_{em} e v_{0xm} / \Omega$,
and the ratio of the E\&M energy intensity to the particle kinetic energy becomes
\begin{equation}
 \frac{\frac{E_1(T)^2}{8\pi}}{ n_{em} m_e(v_{0xm})^2 }
\cong \frac{\alpha^2 }{2}\left(\frac{\omega_{\mathrm{pem}}}{\Omega}\right)^2 \mathrm{,} \label{eq:rat}
\end{equation}
where $ \omega_{\mathrm{pem}}^2 = 4 \pi n_{em} e^2 / m_e $ and $\langle \cos(\Omega t)^2 \rangle = 1/2$
is used.
Eq.~({\ref{eq:rat})
suggests that
the THz E\&M wave gets amplified to the energy intensity comparable to the perpendicular electron
kinetic energy intensity times the ratio $\omega_{\mathrm{pem}}^2 / \Omega^2$
during the time duration of $\Omega$, which  is the maximum energy that could be extracted.
In the moving frame, the perpendicular kinetic energy of an electron  is  $Nm_ev_{0xm}^2 = N\gamma_m^2 m_e v_{0x}^2$.
The maximum total energy radiating into the THz wave is
$E_{\mathrm{max}}= N\gamma_m^2 m_e v_{0x}^2 (\omega_{\mathrm{pem}}^2 / \Omega^2)$
so that  the maximum THz energy in the laboratory frame is
$ \gamma_m E_{\mathrm{max}} =  N\gamma_m^3 m_e v_{0x}^2 (\omega_{\mathrm{pem}}^2 / \Omega^2) $,
where $N$ is the total number of electrons in the beam.
In order to extract the appreciable fraction of the electron kinetic
energy, the ratio  $\omega_{\mathrm{pem}}/ \Omega $ needs to be maximized.
As shown in the non-neutral plasma beam analysis,
it is theoretically possible to get $\omega_{\mathrm{pem}}/ \Omega \cong 1$.

Let us give a few examples of the practically relevant beam parameters. 
Consider a 10 pico-second electron beam 
with $\gamma=35$, $n_e = 10^{13} \ \mathrm{cm^{-3}}$ and
the total number of electrons being $10^{10}$, and 
assume that the magnetic field is order of 1 T. 
If the beam  gets launched with  $v_p / v_{0z} =0.03$, the parallel relativistic factor  is $\gamma_m = 25$.
The resonant frequency for the THz radiation is roughly 1 THz. 
In the moving frame, the electron density becomes roughly
$ 4 \times 10^{11} \  \mathrm{cm^{-3}}$, and $\omega_{\mathrm{pem}}/ \Omega  \cong 0.1 $;
the beam duration is 250 pico-second.
The total energy of the electron is  $7 \times 10^{15} \  \mathrm{eV} $, and
at the maximum, a few percents of the total electron kinetic energy can be radiated into the THz E\&M wave.
As another example, consider a 10 pico-second electron beam with $\gamma=14$.
Assume that the electron density is $10^{14} \ \mathrm{cm^{-3}}$, the total number of electrons is $10^{10}$,
and the beam of $v_p / v_{z0} =0.06$ gets launched ($\gamma_m = 11$).
Assuming the magnetic field is order of 1 T, the resonant frequency is roughly 0.5  THz.
In the moving frame, the electron density is roughly $10^{13} \  \mathrm{cm^{-3}}$,
$\omega_{\mathrm{pem}}/ \Omega  \cong 0.1$, and the beam duration is 100 pico-second.
The total energy of the electron is  $ 10^{16} \ \mathrm{eV} $, and
as much as tens of percents of the total electron kinetic  energy can be radiated into the THz E\&M wave.

To summarize, a scheme of THz generation is discussed,
where the spatial helical structure of the relativistic electron beam is used for
the amplification, via a physical mechanism similar to that of the FEL. 
In contrast to the FEL with the magnets,  
the energy extraction rate from the electrons is not proportional to the intensity,
rather it is proportional to the electric field of the E\&M wave.  
This property makes this scheme advantageous, as the THZ field can be explosively amplified
up to certain amplitude.
The overall efficiency is another advantage.
A THz radiation with the total energy of a few tens of percents  of the total electron beam energy
can be as high as gyrotron or magnetron; the only difference is the operating regime, the THz range.

\bibliography{terahelix}

\begin{thebibliography}{19}
\expandafter\ifx\csname natexlab\endcsname\relax\def\natexlab#1{#1}\fi
\expandafter\ifx\csname bibnamefont\endcsname\relax
  \def\bibnamefont#1{#1}\fi
\expandafter\ifx\csname bibfnamefont\endcsname\relax
  \def\bibfnamefont#1{#1}\fi
\expandafter\ifx\csname citenamefont\endcsname\relax
  \def\citenamefont#1{#1}\fi
\expandafter\ifx\csname url\endcsname\relax
  \def\url#1{\texttt{#1}}\fi
\expandafter\ifx\csname urlprefix\endcsname\relax\def\urlprefix{URL }\fi
\providecommand{\bibinfo}[2]{#2}
\providecommand{\eprint}[2][]{\url{#2}}

\bibitem[{\citenamefont{Nagel et~al.}(2002)\citenamefont{Nagel, Bolivar,
  Brucherseifer, Kurz, Bosserhoff, and Buttner}}]{diagnostic}
\bibinfo{author}{\bibfnamefont{M.}~\bibnamefont{Nagel}},
  \bibinfo{author}{\bibfnamefont{P.~H.} \bibnamefont{Bolivar}},
  \bibinfo{author}{\bibfnamefont{M.}~\bibnamefont{Brucherseifer}},
  \bibinfo{author}{\bibfnamefont{H.}~\bibnamefont{Kurz}},
  \bibinfo{author}{\bibfnamefont{A.}~\bibnamefont{Bosserhoff}},
  \bibnamefont{and} \bibinfo{author}{\bibfnamefont{R.}~\bibnamefont{Buttner}},
  \bibinfo{journal}{Appl.~Phys.~Lett.} \textbf{\bibinfo{volume}{80}},
  \bibinfo{pages}{154} (\bibinfo{year}{2002}).

\bibitem[{\citenamefont{Cooper et~al.}(2008)\citenamefont{Cooper, Dengler,
  Chattopadhyay, Schlecht, Gill, Skalare, Mehdi, and Siegel}}]{radar}
\bibinfo{author}{\bibfnamefont{K.~B.} \bibnamefont{Cooper}},
  \bibinfo{author}{\bibfnamefont{R.~J.} \bibnamefont{Dengler}},
  \bibinfo{author}{\bibfnamefont{G.}~\bibnamefont{Chattopadhyay}},
  \bibinfo{author}{\bibfnamefont{E.}~\bibnamefont{Schlecht}},
  \bibinfo{author}{\bibfnamefont{J.}~\bibnamefont{Gill}},
  \bibinfo{author}{\bibfnamefont{A.}~\bibnamefont{Skalare}},
  \bibinfo{author}{\bibfnamefont{I.}~\bibnamefont{Mehdi}}, \bibnamefont{and}
  \bibinfo{author}{\bibfnamefont{P.~H.} \bibnamefont{Siegel}},
  \bibinfo{journal}{IEEE, Microwave and Wireless Components Letters}
  \textbf{\bibinfo{volume}{18}}, \bibinfo{pages}{64} (\bibinfo{year}{2008}).

\bibitem[{\citenamefont{Siegel}(2007)}]{seigel3}
\bibinfo{author}{\bibfnamefont{P.~H.} \bibnamefont{Siegel}},
  \bibinfo{journal}{Antennas and Propagation, IEEE Transactions on}
  \textbf{\bibinfo{volume}{55}}, \bibinfo{pages}{2957} (\bibinfo{year}{2007}).

\bibitem[{\citenamefont{Yamamoto et~al.}(2004)\citenamefont{Yamamoto,
  Yamaguchi, Miyamaru, Tani, Hangyo, Ikeda, Matsushita, Koide, Tatsuno, and
  Y.Minami}}]{security}
\bibinfo{author}{\bibfnamefont{K.}~\bibnamefont{Yamamoto}},
  \bibinfo{author}{\bibfnamefont{M.}~\bibnamefont{Yamaguchi}},
  \bibinfo{author}{\bibfnamefont{F.}~\bibnamefont{Miyamaru}},
  \bibinfo{author}{\bibfnamefont{M.}~\bibnamefont{Tani}},
  \bibinfo{author}{\bibfnamefont{M.}~\bibnamefont{Hangyo}},
  \bibinfo{author}{\bibfnamefont{T.}~\bibnamefont{Ikeda}},
  \bibinfo{author}{\bibfnamefont{A.}~\bibnamefont{Matsushita}},
  \bibinfo{author}{\bibfnamefont{K.}~\bibnamefont{Koide}},
  \bibinfo{author}{\bibfnamefont{M.}~\bibnamefont{Tatsuno}}, \bibnamefont{and}
  \bibinfo{author}{\bibnamefont{Y.Minami}},
  \bibinfo{journal}{Jpn.~J.~Appl.~Phys.} \textbf{\bibinfo{volume}{43}},
  \bibinfo{pages}{L414} (\bibinfo{year}{2004}).

\bibitem[{\citenamefont{Siegel}(2002)}]{siegel}
\bibinfo{author}{\bibfnamefont{P.~H.} \bibnamefont{Siegel}},
  \bibinfo{journal}{Microwave Theory and Techniques, IEEE Transaction on}
  \textbf{\bibinfo{volume}{50}}, \bibinfo{pages}{910} (\bibinfo{year}{2002}).

\bibitem[{\citenamefont{Siegel}(2004)}]{siegel2}
\bibinfo{author}{\bibfnamefont{P.~H.} \bibnamefont{Siegel}},
  \bibinfo{journal}{Microwave Theory and Techniques, IEEE Transaction on}
  \textbf{\bibinfo{volume}{52}}, \bibinfo{pages}{2438} (\bibinfo{year}{2004}).

\bibitem[{\citenamefont{Chu et~al.}(1998)\citenamefont{Chu, Chen, Hung, Chang,
  Barnett, Chen, and Yang}}]{gyrotron}
\bibinfo{author}{\bibfnamefont{K.~R.} \bibnamefont{Chu}},
  \bibinfo{author}{\bibfnamefont{H.~Y.} \bibnamefont{Chen}},
  \bibinfo{author}{\bibfnamefont{C.~L.} \bibnamefont{Hung}},
  \bibinfo{author}{\bibfnamefont{T.~H.} \bibnamefont{Chang}},
  \bibinfo{author}{\bibfnamefont{L.~R.} \bibnamefont{Barnett}},
  \bibinfo{author}{\bibfnamefont{S.~H.} \bibnamefont{Chen}}, \bibnamefont{and}
  \bibinfo{author}{\bibfnamefont{T.~T.} \bibnamefont{Yang}},
  \bibinfo{journal}{Phys.~Rev.~Lett.} \textbf{\bibinfo{volume}{81}},
  \bibinfo{pages}{4760} (\bibinfo{year}{1998}).

\bibitem[{\citenamefont{Kreischer and Temkin}(547)}]{gyrotron2}
\bibinfo{author}{\bibfnamefont{K.~E.} \bibnamefont{Kreischer}}
  \bibnamefont{and} \bibinfo{author}{\bibfnamefont{R.~J.}
  \bibnamefont{Temkin}}, \bibinfo{journal}{Phys.~Rev.~Lett.}
  \textbf{\bibinfo{volume}{59}}, \bibinfo{pages}{1987} (\bibinfo{year}{547}).

\bibitem[{\citenamefont{Bratman et~al.}(2009)\citenamefont{Bratman, Kalynov,
  and Manuilov}}]{gyrotron3}
\bibinfo{author}{\bibfnamefont{V.~L.} \bibnamefont{Bratman}},
  \bibinfo{author}{\bibfnamefont{Y.~L.} \bibnamefont{Kalynov}},
  \bibnamefont{and} \bibinfo{author}{\bibfnamefont{V.~N.}
  \bibnamefont{Manuilov}}, \bibinfo{journal}{Phys.~Rev.~Lett.}
  \textbf{\bibinfo{volume}{102}}, \bibinfo{pages}{245101}
  (\bibinfo{year}{2009}).

\bibitem[{\citenamefont{Booske}(2008)}]{booske}
\bibinfo{author}{\bibfnamefont{J.~H.} \bibnamefont{Booske}},
  \bibinfo{journal}{Physics of Plasmas} \textbf{\bibinfo{volume}{15}},
  \bibinfo{pages}{055502} (\bibinfo{year}{2008}).

\bibitem[{\citenamefont{Faist et~al.}(1994)\citenamefont{Faist, Capasso, Sivco,
  Sirtori, Hutchinson, and Cho}}]{qlaser}
\bibinfo{author}{\bibfnamefont{J.}~\bibnamefont{Faist}},
  \bibinfo{author}{\bibfnamefont{F.}~\bibnamefont{Capasso}},
  \bibinfo{author}{\bibfnamefont{D.~L.} \bibnamefont{Sivco}},
  \bibinfo{author}{\bibfnamefont{C.}~\bibnamefont{Sirtori}},
  \bibinfo{author}{\bibfnamefont{A.~L.} \bibnamefont{Hutchinson}},
  \bibnamefont{and} \bibinfo{author}{\bibfnamefont{A.~Y.} \bibnamefont{Cho}},
  \bibinfo{journal}{Science} \textbf{\bibinfo{volume}{264}},
  \bibinfo{pages}{553} (\bibinfo{year}{1994}).

\bibitem[{\citenamefont{Tonouchi}(2007)}]{qlaser2}
\bibinfo{author}{\bibfnamefont{M.}~\bibnamefont{Tonouchi}},
  \bibinfo{journal}{Nature Photonics} \textbf{\bibinfo{volume}{1}},
  \bibinfo{pages}{97} (\bibinfo{year}{2007}).

\bibitem[{\citenamefont{Colson}(1985)}]{colson}
\bibinfo{author}{\bibfnamefont{W.~B.} \bibnamefont{Colson}},
  \bibinfo{journal}{Nucl.~Inst.~Meth.~Phys.} \textbf{\bibinfo{volume}{A237}},
  \bibinfo{pages}{1} (\bibinfo{year}{1985}).

\bibitem[{\citenamefont{Malkin and Fisch}(2007)}]{Fisch}
\bibinfo{author}{\bibfnamefont{V.~M.} \bibnamefont{Malkin}} \bibnamefont{and}
  \bibinfo{author}{\bibfnamefont{N.~J.} \bibnamefont{Fisch}},
  \bibinfo{journal}{Phys.~Rev.~Lett.} \textbf{\bibinfo{volume}{99}},
  \bibinfo{pages}{205001} (\bibinfo{year}{2007}).

\bibitem[{\citenamefont{Son and Ku}(2009)}]{sonlandau}
\bibinfo{author}{\bibfnamefont{S.}~\bibnamefont{Son}} \bibnamefont{and}
  \bibinfo{author}{\bibfnamefont{S.}~\bibnamefont{Ku}},
  \bibinfo{journal}{Phys.~Plasmas} \textbf{\bibinfo{volume}{17}},
  \bibinfo{pages}{010703} (\bibinfo{year}{2009}).

\bibitem[{\citenamefont{Son et~al.}(2010)\citenamefont{Son, Ku, and
  Moon}}]{sonbackward}
\bibinfo{author}{\bibfnamefont{S.}~\bibnamefont{Son}},
  \bibinfo{author}{\bibfnamefont{S.}~\bibnamefont{Ku}}, \bibnamefont{and}
  \bibinfo{author}{\bibfnamefont{S.~J.} \bibnamefont{Moon}},
  \bibinfo{journal}{Phys.~Plasmas} \textbf{\bibinfo{volume}{17}},
  \bibinfo{pages}{114506} (\bibinfo{year}{2010}).

\bibitem[{\citenamefont{Tabak et~al.}(1994)\citenamefont{Tabak, Hammer,
  Glinsky, Kruerand, Wilks, Woodworth, Campbell, Perry, and Mason}}]{tabak}
\bibinfo{author}{\bibfnamefont{M.}~\bibnamefont{Tabak}},
  \bibinfo{author}{\bibfnamefont{J.}~\bibnamefont{Hammer}},
  \bibinfo{author}{\bibfnamefont{M.~E.} \bibnamefont{Glinsky}},
  \bibinfo{author}{\bibfnamefont{W.~L.} \bibnamefont{Kruerand}},
  \bibinfo{author}{\bibfnamefont{S.~C.} \bibnamefont{Wilks}},
  \bibinfo{author}{\bibfnamefont{J.}~\bibnamefont{Woodworth}},
  \bibinfo{author}{\bibfnamefont{E.~M.} \bibnamefont{Campbell}},
  \bibinfo{author}{\bibfnamefont{M.~J.} \bibnamefont{Perry}}, \bibnamefont{and}
  \bibinfo{author}{\bibfnamefont{R.~J.} \bibnamefont{Mason}},
  \bibinfo{journal}{Physics of Plasmas} \textbf{\bibinfo{volume}{1}},
  \bibinfo{pages}{1626} (\bibinfo{year}{1994}).

\bibitem[{\citenamefont{Son and Fisch}(2004)}]{sonpla}
\bibinfo{author}{\bibfnamefont{S.}~\bibnamefont{Son}} \bibnamefont{and}
  \bibinfo{author}{\bibfnamefont{N.~J.} \bibnamefont{Fisch}},
  \bibinfo{journal}{Phys.~Lett.~A} \textbf{\bibinfo{volume}{329}},
  \bibinfo{pages}{16} (\bibinfo{year}{2004}).

\bibitem[{\citenamefont{Son and Fisch}(2005)}]{sonprl}
\bibinfo{author}{\bibfnamefont{S.}~\bibnamefont{Son}} \bibnamefont{and}
  \bibinfo{author}{\bibfnamefont{N.~J.} \bibnamefont{Fisch}},
  \bibinfo{journal}{Phys.~Rev.~Lett.} \textbf{\bibinfo{volume}{95}},
  \bibinfo{pages}{225002} (\bibinfo{year}{2005}).

\end{thebibliography}

\end{document}